\documentclass[]{aa}
\usepackage{graphicx}
\usepackage{natbib}

\usepackage{color}

\usepackage{amssymb}

\usepackage[varg]{txfonts}

\usepackage{amstext}

\begin{document}

\definecolor{bleu}{rgb}{.3,0.2,1.0}
\definecolor{red}{rgb}{1.,0.2,0.2}

\title{The magnetic nature of umbra-penumbra boundary in sunspots }

\author{J. Jur\v{c}\'{a}k
        \inst{1}
        \and
        R. Rezaei
        \inst{2, 3}
        \and
        N. Bello Gonz\'{a}lez
        \inst{4}
        \and
        R. Schlichenmaier
        \inst{4}
        \and
        J. Vomlel
        \inst{5}}

\institute{Astronomical Institute of the Academy of Sciences, Fri\v{c}ova  298, 25165 Ond\v{r}ejov, Czech Republic
  \and
  Instituto de Astrof\'isica de Canarias (IAC), V\'ia Lact\'ea, 38200 La Laguna (Tenerife), Spain 
  \and
  Departamento de Astrof\'isica, Universidad de La Laguna, 38205 La Laguna (Tenerife), Spain
  \and
  Kiepenheuer-Institut f\"{u}r Sonnenphysik, Sch\"{o}neckstr. 6, 79104 Freiburg, Germany
  \and
  The Czech Academy of Sciences, Institute of Information Theory and Automation, Pod vod\'{a}renskou v\v{e}\v{z}\'{\i} 4, 182 08 Praha, Czech Republic}

\date{Received 11 December, 2014; accepted }

\abstract
{Sunspots are the longest-known manifestation of solar activity, and their magnetic nature has been known for more than a century. Despite this, the boundary between umbrae and penumbrae, the two fundamental sunspot regions, has hitherto been solely defined by an intensity threshold.}
  {Here, we aim at studying the magnetic nature of umbra-penumbra boundaries in sunspots of different sizes, morphologies, evolutionary stages, and phases of the solar cycle.}
 {We used a sample of 88 scans of the Hinode/SOT spectropolarimeter to infer the magnetic field properties in at the umbral boundaries. We defined these umbra-penumbra boundaries by an intensity threshold and performed a statistical analysis of the magnetic field properties on these boundaries.}
{We statistically prove that the umbra-penumbra boundary in stable sunspots is characterised by an invariant value of the vertical magnetic field component: the vertical component of the magnetic field strength does not depend on the umbra size, its morphology, and phase of the solar cycle. With the statistical Bayesian inference, we find that the strength of the vertical magnetic field component is, with a likelihood of 99\%, in the range of 1849-1885 G with the most probable value of 1867 G. In contrast, the magnetic field strength and inclination averaged along individual boundaries are found to be dependent on the umbral size: the larger the
umbra, the stronger and more horizontal the magnetic field at its boundary.}
{The umbra and penumbra of sunspots are separated by a boundary that has hitherto been defined by an intensity threshold. We now unveil the empirical law of the magnetic nature of the umbra-penumbra boundary in stable sunspots: it is an invariant vertical component of the magnetic field.}
  

\maketitle

%
%

\section{Introduction}
\label{introduction}

The enhanced temperature and brightness of a penumbra compared to temperature and brightness of an umbra define a sharp intensity boundary between these regions. This umbral boundary is commonly outlined by an intensity threshold of 50\% relative to the spatially averaged surrounding quiet-Sun intensity in visible continuum. An intensity threshold was necessary to define the umbral boundary because the magnetic nature of this boundary was unknown. 

The magnetic nature of sunspots was discovered by \citet{Hale:1908}. Several analyses have described the global properties of the magnetic field in sunspots \citep{Lites:1990, Solanki:1992, Balthasar:1993, Keppens:1996, cwp:2001, Mathew:2003, Borrero:2004, Bellot:2004, Balthasar:2005,   Sanchez:2005, Beck:2008, Borrero:2011}. The umbra harbours stronger and more vertical magnetic field than the penumbra. With increasing radial distance from the umbral core, the fields becomes weaker and more horizontal. Detailed analyses of penumbral filaments discovered that the horizontal fields are the essential property of these structures \citep{Tiwari:2013, Jurcak:2014}. These horizontal fields are interlaced with a background component of the penumbral magnetic field creating the so-called \textup{uncombed }structure of sunspot penumbrae \citep{Solanki:1993}.

Despite the detailed knowledge of the magnetic field structure, the analyses did not point to any specific properties of the magnetic field at the umbra/penumbra (UP) boundary. Reasons for this inconclusiveness are discussed in \citet{Jurcak:2011}. In this analysis, the magnetic field strength and inclination were investigated directly at the UP boundaries. \citet{Jurcak:2011} found evidence that the UP boundaries are defined by a constant value of the vertical component of the magnetic field, $B_\textrm{ver}$. Although it changes smoothly across the boundary, $B_\textrm{ver}$ is found to be constant along the boundary. In contrast, the magnetic field strength and inclination vary along the boundary. The follow-up case study showed that in a forming spot, umbral areas with $B_\textrm{ver}$ lower than this constant value are colonised by the penumbra \citep{Jurcak:2015}. In addition, an umbra with overall reduced $B_\textrm{ver}$ was observed to fully transform into a penumbra giving birth to a so-called orphan penumbra \citep{Jurcak:2017}. 

\begin{figure*}[!t]
 \centering \includegraphics[width=0.95\linewidth]{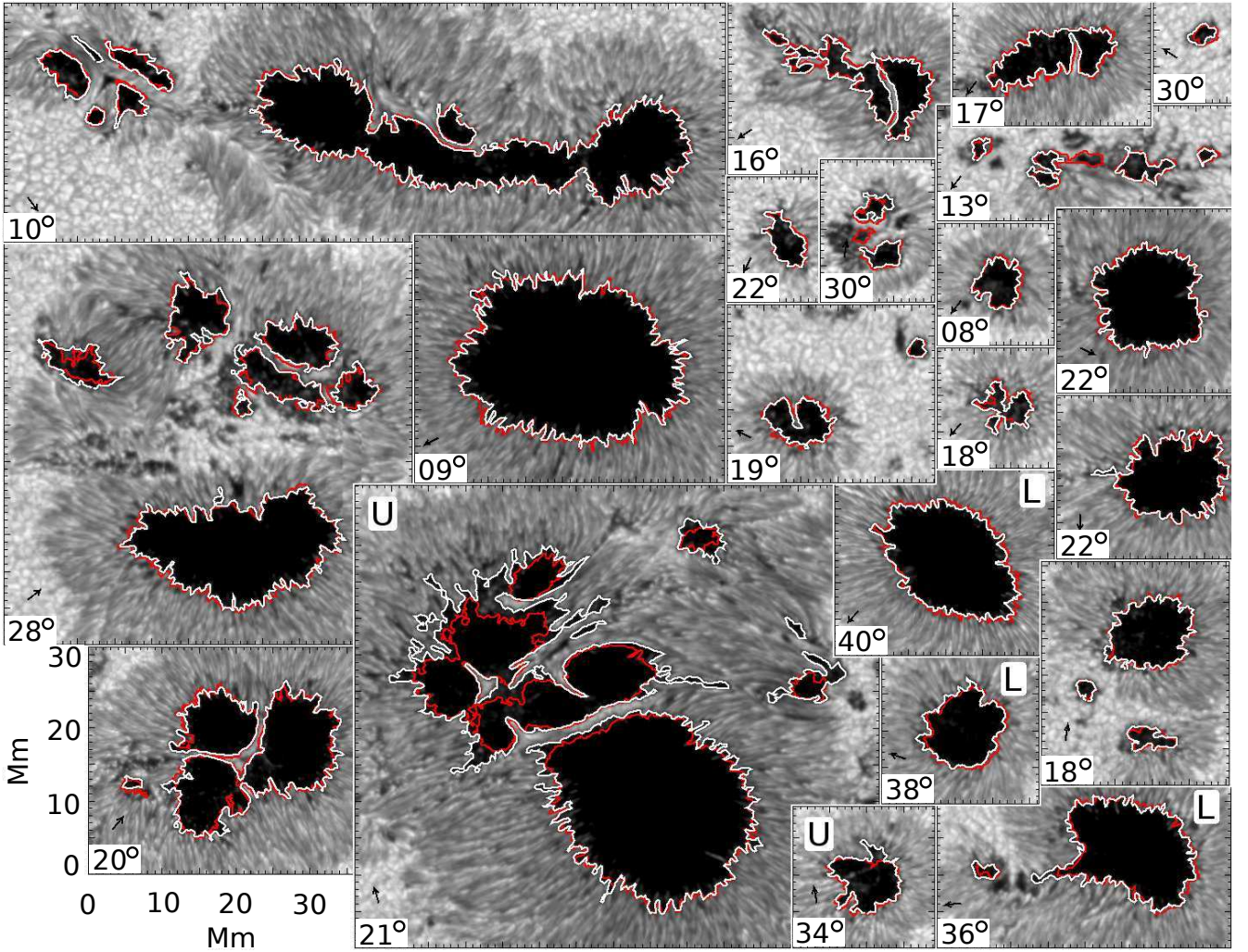}
 \caption{Selection of the analysed sample of sunspots displayed on the same scale. The white contours mark the intensity threshold of 50\% of the quiet-Sun intensity. The red contours are independently defined and outline isocontours of 1867~G of $B_\textrm{ver}$. Only contours encircling regions larger than 3~Mm$^2$ are shown. Sunspots marked (L) clearly show a systematic displacement of the white and red contours that is due to line-of-sight effects. Sunspots marked (U) have regions where the red contours lie within the intensity (white) contours. The arrows point to the disc centre. The numbers denote the heliocentric angle of the sunspot.}
 \label{sample}
\end{figure*}

In this study, we carry out a statistical analysis of the magnetic field properties on boundaries of more than 100 umbral cores. We compare our results with the proposed canonical value of $B_\textrm{ver}$ to check for a proof of its invariance. We discuss the sample of sunspots and the data analysis in Sect.~\ref{observations} and present the properties of the UP boundaries in  Sect.~\ref{results}. We discuss the results and conclude in Sect.~\ref{discussion}.

\section{Observations and data analysis}
\label{observations}

We determined the magnetic field vector using 88 scans of 79 different active regions observed with the spectropolarimeter (SP) attached to the Solar Optical Telescope \citep{Tsuneta:2008} on board the Hinode satellite \citep{Kosugi:2007} from 2006 to 2015, in the course of solar cycle 24. The sample contains sunspots of different sizes, morphologies, evolutionary stages, and phases of the solar cycle. A full list of the observed regions is given in Appendix~A. 

Hinode SP records full Stokes profiles of the neutral iron line pair at 630 nm. The observed line profiles were calibrated using the standard reduction routines \citep{Lites:2013}. The majority of the Hinode SP scans are taken in so-called fast mode, for which the spatial sampling is 0\farcs32. The SP scans taken with a spatial sampling of 0\farcs16 were smoothed by a boxcar function with the size $2\times2$ pixels to minimize the effect of different spatial resolution in our sample, ensuring the homogeneity of our dataset. For each scan, we constructed continuum intensity maps using the Stokes $I$ profile intensities around 630.1 nm. We also determined the quiet-Sun intensity for each scan and computed isocontours at 50\% of this intensity. We derived areas encircled by these contours and corrected them for projection effects. For the further analysis, we considered only isocontours encircling areas larger than 3~Mm$^2$. 

To determine the magnetic field properties, we performed SIR inversion \citep[Stokes inversion based on response function,][]{Cobo:1992}. Except for the temperature, all atmospheric parameters were considered height independent. We took into account the spectral point-spread function of the Hinode SP in the inversion process. We assumed the magnetic filling factor to be unity and assumed no stray light. The macroturbulence was set to zero, while microturbulence was a free parameter of the inversion. The retrieved magnetic field vectors were then transformed into the local reference frame, and the azimuth ambiguity was removed using the AZAM module \citep{Lites:1995}. The 0\,deg inclination angle ($\gamma$) defines the vertical direction, that is, the direction perpendicular to the solar surface. We inverted the polarities of negative sunspots. 

The averaged magnetic field strength ($B$), inclination ($\gamma$), and vertical component ($B_\textrm{ver}$) along umbra boundaries were then computed along each isocontour defined by 50\% of the mean quiet-Sun intensity for each umbral core. 

The homogeneity of the dataset provided by Hinode, free from seeing effects and with an identical instrument setup for all sunspot maps, has allowed us to statistically analyse the magnetic parameters on the umbral boundaries, that is, to investigated the dependence of $B$, $\gamma$, and $B_\textrm{ver}$ (dependent variables) on the logarithm of the area encircled by the intensity isocontours and also the dependence of $B_\textrm{ver}$ on the date of the observations (explanatory variables). We primarily used a Bayesian linear regression. To support our conclusions, we also used standard linear regression. Both methods lead to the same conclusions, which are described in detail in Appendix~B.

\section{Results}
\label{results}

\begin{figure}[!t]
 \centering \includegraphics[width=\linewidth]{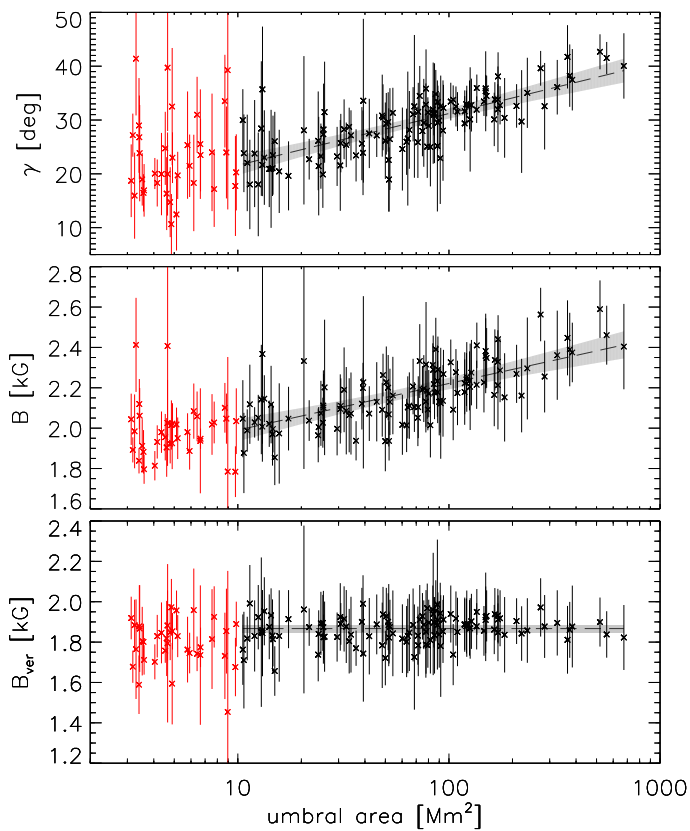}
 \caption{Mean values of the magnetic field inclination (A), the magnetic field strength (B), and the vertical component of the magnetic field (C) for every identified boundary  as a function of the area encircled by the given boundary. The uncertainties represent the standard deviations of the physical parameters at the boundaries. Contours encircling areas smaller than 10~Mm$^2$ marked in red are not considered for the statistical analyses. The dashed line corresponds to the most probable value estimated by a Bayesian linear regression, and the shaded area denotes the 99\% confidence interval of the estimated model.} 
 \label{dep_area}
\end{figure}

\begin{figure}[!t]
 \sidecaption
 \includegraphics[width=\linewidth]{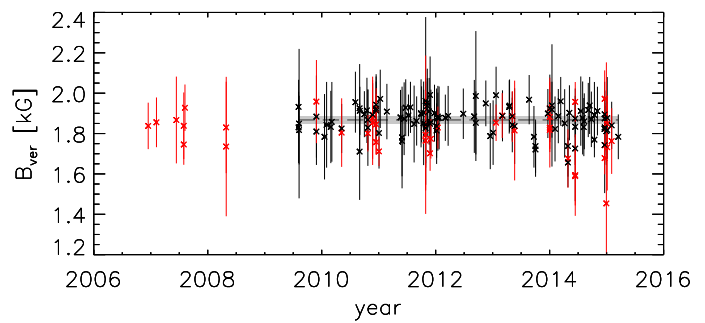}
 \caption{Dependence of $B_\textrm{ver}$ on the phase of the solar cycle. Contours encircling areas smaller than 10~Mm$^2$ and sunspots observed in solar cycle 23 are marked in red and not considered for the statistical analyses. The dashed line corresponds to the most probable value estimated by a Bayesian linear regression, and the shaded area denotes the 99\% confidence interval of the estimated model.} 
 \label{dep_time} 
\end{figure}

Figure~\ref{sample} presents a selection of the analysed sample of sunspots in which the striking correspondence between the independently defined boundaries based on thresholds of intensity (white contours) and $B_\textrm{ver} = 1867$~G (red contours) manifests the fundamental invariance of $B_\textrm{ver}$. In no single case does the contour of $B_\textrm{ver}$ cross significantly into the penumbra. There is a small systematic line-of-sight effect for sunspots observed far-off disk centre: since the absorption line to infer $B_\textrm{ver}$ forms some 200 km higher than the continuum intensity, the contours for intensity and $B_\textrm{ver}$ are shifted correspondingly for sunspots observed close to the limb. This effect is recognisable in sunspots marked L in Figure 1. In some sunspots, the intensity boundary is not coupled with the $B_\textrm{ver}$ contour (examples marked U in Fig.~\ref{sample}), but the latter crosses through umbral areas. These cases are discussed in Sect.~\ref{discussion}. Our sample of sunspots also exhibits light bridges dividing umbrae into multiple umbral cores. We find that the invariant magnetic property equally holds at the boundaries of light bridges.

Mean values of $B$, $\gamma$, and $B_\textrm{ver}$ were computed along the intensity boundaries of all umbral regions larger than 3~Mm$^2$ for each sunspot in the sample. Figure~\ref{dep_area} shows the dependence on the umbral area of the mean values of $\gamma$, $B$, and $B_\textrm{ver}$ along the intensity isocontour. The changes in magnetic field inclination and strength (panels A and B) are correlated: the larger the umbra, the stronger and more horizontal the magnetic field at its boundary. The vertical component, $B_\textrm{ver}$ (panel C), does not depend on the size of the umbral area. These findings are corroborated by a statistical Bayesian inference applied to our dataset. With this analysis, we find that $B_\textrm{ver}$ is most likely independent of the umbral size and is in the range of $1849-1885$~G with a likelihood of 99\%; the most probable value is 1867~G. 

Since the maximum magnetic field strength in sunspots shows a systematic variation with the phase of the solar cycle \citep{Rezaei:2012a, Pevtsov:2014, Schad:2014, Rezaei:2015}, we investigated the dependence of the $B_\textrm{ver}$ with time throughout solar cycle 24 (Fig.~\ref{dep_time}). The Bayesian inference method identifies the constant model as the most plausible to describe this dependence. The parameters of the constant fit are the same as for the dependence of $B_\textrm{ver}$ on the umbra area. 

We have thus unveiled the empirical law governing the boundary between the modes of energy transport operating in the umbra and penumbra of stable sunspots: the vertical component of the magnetic field strength discriminates the umbral from the penumbral mode of magneto-convection and an invariant value of $B_\textrm{ver}$ allows identifying the boundary between umbra and penumbra in stable sunspots, reflecting its magnetic nature. The value of this invariant is subjected to the measurement process. With our data, inversion method, and statistical analysis, we find a most probable value of 1867~G.

\section{Discussion and conclusions}
\label{discussion}

While in stable sunspots the intensity and the $B_\textrm{ver}$ boundary contours are coupled, cases are also observed in which the $B_\textrm{ver}=1867$~G isocontour lies within the intensity contour \citep[see sunspots marked U in Figure 1 and][]{Jurcak:2015, Jurcak:2017}. We conjecture that the empirical law presented here serves as a stability criterion in evolving spots and proto-spots: pores and umbral areas with $B_\textrm{ver}$ higher than 1867~G are stable, while umbral areas with lower $B_\textrm{ver}$ are unstable. That is to say, they are prone to be transformed into a different mode of magneto-convection like a penumbra or light bridges. This is also why we restricted our statistical analysis to umbral areas larger than 10~Mm$^2$. Smaller areas are typically observed to be in a transient state.

Moreover, light bridges harbour horizontal magnetic fields \citep{Beckers:1969, Jurcak:2006, Felipe:2016}. We find that
our empirical law indeed applies here as well, as demonstrated in the examples shown in Fig.~\ref{sample}. Based on images of very high spatial resolution, \citet{Schlichenmaier:2016} reported that the fine structure in the boundary of light bridges is found to share morphological aspects with the inner boundary of penumbral filaments, suggesting that the modes of magneto-convection in light bridges and penumbrae are coupled to the umbral mode in a similar manner. 

The implications of these findings have profound consequences on the formation process of the penumbra. While the umbra harbours strong vertical fields \citep{Bray:1964, Solanki:2003}, horizontal fields are the essential property of penumbral filaments \citep{Tiwari:2013, Jurcak:2014}. During the formation of the penumbra, the umbral magnetic field with sufficiently low $B_\textrm{ver}$ is converted into a penumbral magnetic field \citep{Jurcak:2015, Jurcak:2017}.

There must therefore exist a mechanism that turns vertical (umbral) fields into horizontal (penumbral) fields. An explanation of this process regulated by the interplay between the convective and magnetic forces seems straightforward, but is fundamentally new: from the convectively unstable sub-photosphere, hot buoyant plasma pushes upwards along the umbral field lines. When cooling, the mass load exerted by the plasma will bend and incline the field lines of a reduced (lower than 1867~G) vertical field. Only in strong vertical magnetic fields (higher than 1867~G)
is the magnetic tension as a restoring force strong enough to prevent the field lines from bending over and becoming horizontal. This scenario is consistent with the fallen flux tube model by \citet{Wentzel:1992} and with numerical simulations of the penumbral fine structure, which showed that the mass load in inclined penumbral field lines overcomes the magnetic tension to form a horizontal penumbral filament \citep{Rempel:2011}.

The discovery of the empirical law defining the existence of a critical vertical component of the magnetic field governing the umbra boundary in stable sunspots solves the long-standing question on the nature of this boundary, and it gives fundamental new insights into the magneto-convective modes of energy transport in sunspots, which will be addressed in following studies.

\begin{acknowledgements}

The authors wish to thank I. Thaler for fruitful discussions and A. Asensio Ramos for a preliminary Bayesian analysis of the data. J.J. acknowledges the support from RVO:67985815. R.R. acknowledges financial support by the DFG grant RE 3282/1-1 and by the Spanish Ministry of Economy and Competitiveness through project AYA2014-60476-P. N.B.G. acknowledges financial support by the Senatsausschuss of the Leibniz-Gemeinschaft, Ref.-No. SAW-2012-KIS-5 within the framework of the CASSDA project. Hinode is a Japanese mission developed and launched by ISAS/JAXA, with NAOJ as domestic partner and NASA and STFC (UK) as international partners. It is operated by these agencies in cooperation with ESA and NSC (Norway). 

\end{acknowledgements}

\bibliographystyle{aa}
\bibliography{manuscript}

\begin{appendix}
 
 \section{Datasets}
 
\begin{figure*}[!t]
 \centering \includegraphics[width=0.94\linewidth]{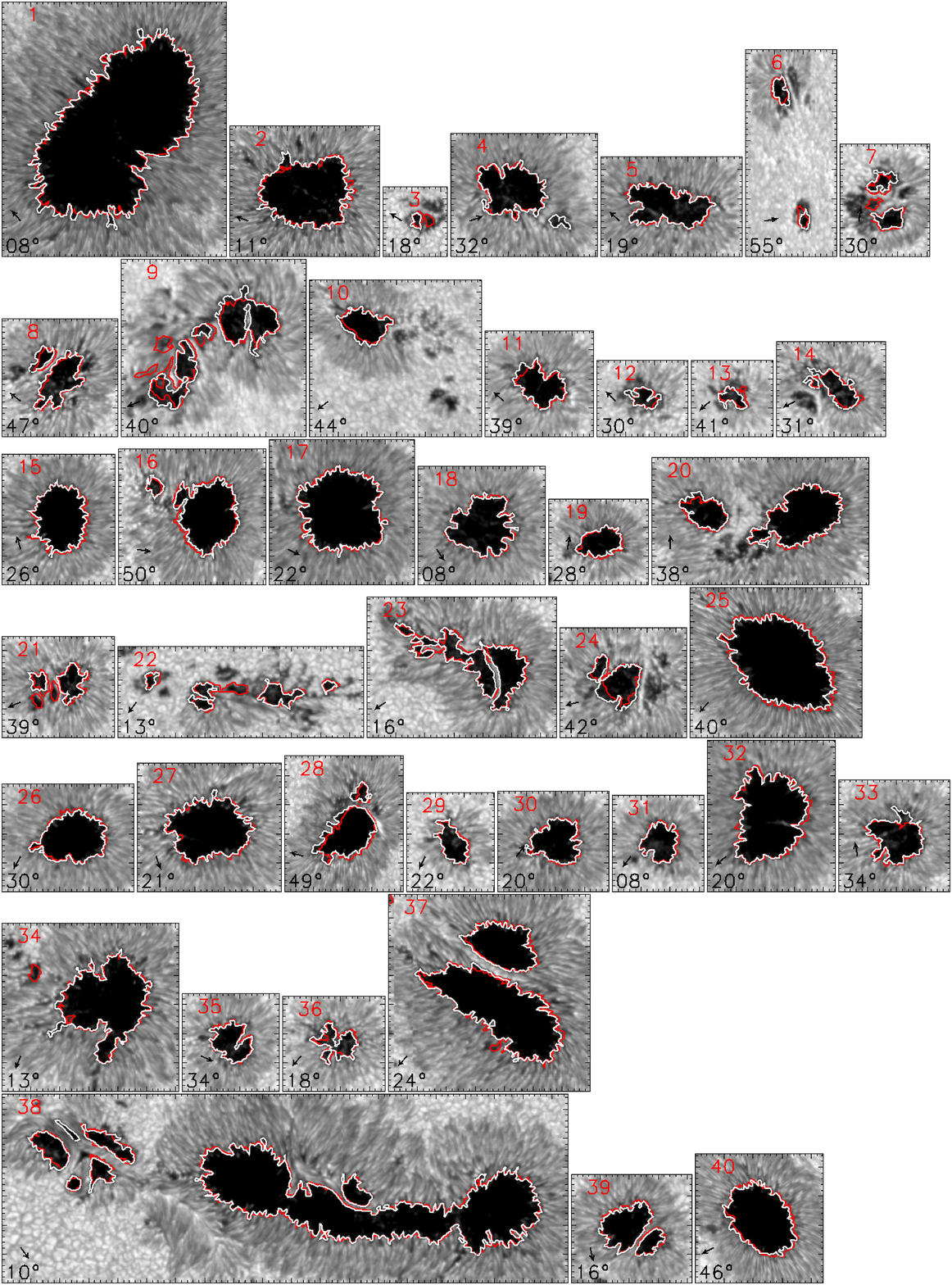}
 \caption{Hinode SP maps. The spatial scale in all panels is the same, with one tick mark of the axis being 1\arcsec. The white contours mark the intensity threshold of 50\% of the quiet-Sun intensity. The red contours are independently defined and outline the isocontours of 1867~G of $B_\textrm{ver}$. Only contours encircling regions larger than 3~Mm$^2$ are shown. The arrows point to the disc centre. The black numbers denote the heliocentric angle of the sunspot. The red numbers refer to the scan number as listed in Table~\ref{table_data}.}
 \label{fig_data1}
 \end{figure*}

 In Table~\ref{table_data} we summarise all the datasets that were used for the statistical analysis. For each listed item in Table~\ref{table_data}, we constructed a continuum intensity map with marked intensity isocontours at 50\% of the local quiet-Sun intensity and isocontours at $B_\textrm{ver} = 1867$~G. Example of these maps is shown in Fig.~\ref{fig_data1} (the rest not display for file-size limits on arxiv). When the date, time, and NOAA numbers were identical in Table~\ref{table_data}, the scan was separated into sub-fields for display purposes.
 
 \begin{table*}
\caption{List of Hinode SP scans.}
\label{table_data}
\centering 
\begin{tabular}{c c c c | c c c c }
Scan & Date & Time (UT) & NOAA & Scan & Date & Time (UT) & NOAA \\ 
\hline 
1 & 2006/11/14 & 16:30 & 10923 & 48 & 2012/02/19 & 19:00 & 11420 \\
2 & 2007/01/06  & 13:00 & 10933 & 49 & 2012/05/25 & 07:44 & 11486 \\
3 & 2007/05/12 & 11:43 & 10955 & 50 & 2012/08/04 & 19:07 & 11538 \\
4 & 2007/06/29 & 09:13 & 10961 & 51 & 2012/08/14 & 00:00 & 11543\\
5 & 2007/07/02 & 12:38 & 10961 & 52 & 2012/10/17 & 09:06 & 11591\\
6 & 2008/03/27 & 11:50 & 10989 & 53 & 2012/11/15 & 20:18 & 11613\\
7 & 2009/07/04 & 12:18 & 11024 & 54 & 2012/12/03 & 18:52 & 11625 \\
8 & 2009/07/07 & 10:45 & 11024 & 55 & 2012/12/22 & 02:46 & 11633 \\
9 & 2009/10/27 & 10:45 & 11029 & 56 & 2013/01/31 & 13:06 & 11663 \\
10 & 2009/12/18 & 14:07 & 11035 & 57 & 2013/02/02 & 07:06 & 11665 \\
11 & 2010/01/01 & 11:50 & 11039 & 58 & 2013/03/15 & 09:44 & 11692 \\
12 & 2010/01/28 & 11:15 & 11041 & 59 & 2013/03/17 & 11:50 & 11692 \\
13 & 2010/02/04 & 17:00 & 11043 & 60 & 2013/04/04 & 15:50 & 11711 \\
14 & 2010/04/06 & 22:00 & 11061 & 61 & 2013/04/19 & 15:44 & 11723 \\
  
15 & 2010/07/03 & 00:32 & 11084 & 62 & 2013/07/21 & 13:18 & 11793 \\
  
16 & 2010/07/30 & 17:32 & 11092 & 63 & 2013/08/21 & 22:04 & 11823 \\
  
17 & 2010/08/01 & 20:30 & 11092 & 64 & 2013/09/01 & 13:05 & 11836 \\
  
18 & 2010/08/30 & 02:23 & 11101 & 65 & 2013/11/19 & 11:02 & 11899 \\
  
19 & 2010/09/16 & 16:30 & 11106 & 66 & 2013/12/01 & 10:00 & 11908 \\
  
20 & 2010/09/22 & 09:30 & 11108 & 67 & 2013/12/15 & 02:05 & 11921 \\
  
21 & 2010/10/22 & 20:40 & 11113 & 68 & 2014/01/04 & 19:26 & 11944 \\
  
22 & 2010/11/14 & 15:36 & 11124 & 69 & 2014/01/25 & 06:59 & 11959 \\
  
23 & 2010/11/14 & 15:36 & 11124 & 70 & 2014/02/10 & 08:13 & 11974 \\
  
24 & 2010/12/02 & 01:37 & 11130 & 71 & 2014/03/05 & 23:24 & 11990 \\
  
25 & 2010/12/10 & 16:24 & 11131 & 72 & 2014/03/26 & 15:20 & 12014 \\
  
26 & 2011/01/22 & 09:49 & 11147 & 73 & 2014/03/26 & 15:20 & 12014 \\
  
27 & 2011/04/18 & 16:47 & 11193 & 74 & 2014/04/05 & 13:24 & 12027 \\
  
28 & 2011/04/28 & 13:10 & 11195 & 75 & 2014/05/12 & 02:15 & 12056 \\
  
29 & 2011/05/11 & 01:08 & 11210 & 76 & 2014/05/12 & 02:15 & 12056 \\
  
30 & 2011/05/21 & 03:43 & 11216 & 77 & 2014/05/12 & 02:15 & 12056 \\
  
31 & 2011/06/08 & 00:32 & 11232 & 78 & 2014/05/12 & 02:15 & 12056 \\
  
32 & 2011/06/21 & 11:54 & 11236 & 79 & 2014/06/17 & 22:55 & 12087 \\
  
33 & 2011/07/14 & 19:33 & 11250 & 80 & 2014/06/29 & 05:29 & 12096 \\
  
34 & 2011/07/30 & 14:28 & 11260 & 81 & 2014/07/06 & 01:19 & 12104 \\
  
35 & 2011/08/28 & 09:54 & 11277 & 82 & 2014/07/09 & 05:00 & 12109 \\
  
36 & 2011/09/01 & 07:52 & 11277 & 83 & 2014/08/12 & 18:15 & 12135 \\
  
37 & 2011/09/14 & 14:55 & 11289 & 84 & 2014/08/13 & 01:21 & 12135 \\
  
38 & 2011/09/27 & 18:35 & 11302 & 85 & 2014/08/24 & 18:18 & 12146 \\
  
39 & 2011/10/07 & 22:30 & 11309 & 86 & 2014/09/13 & 02:23 & 12158 \\
  
40 & 2011/10/19 & 06:42 & 11314 & 87 & 2014/09/13 & 02:23 & 12158 \\
  
41 & 2011/10/25 & 14:36 & 11330 & 88 & 2014/10/02 & 14:19 & 12181 \\
  
42 & 2011/10/30 & 17:30 & 11330 & 89 & 2014/11/18 & 19:43 & 12209 \\
  
43 & 2011/11/23 & 00:15 & 11352 & 90 & 2014/11/29 & 16:40 & 12222 \\
  
44 & 2011/12/06 & 23:08 & 11363 & 91 & 2014/12/04 & 01:00 & 12222 \\
  
45 & 2011/12/14 & 00:34 & 11374 & 92 & 2014/12/04 & 04:00 & 12222 \\
  
46 & 2011/12/24 & 04:21 & 11384 & 93 & 2015/01/02 & 08:30 & 12251 \\
  
47 & 2012/02/01 & 19:15 & 11410 & 94 & 2015/02/14 & 15:00 & 12282 \\
\hline 
\end{tabular}
\end{table*}
 
 \section{Statistical analysis}
 
We performed both a Bayesian linear regression and a standard linear regression.  Before we performed these statistical analyses, we centred and scaled the $x$ values of each explanatory variable to ensure
better numerical stability: $$
 x'=\frac{x-\mu(x)}{2\sigma(x)},
$$
where $\mu(x)$ is the mean of the explanatory variable, and $\sigma(x)$ is the standard deviation of the explanatory variable.
 
 \subsection{Bayesian linear regression}
 
 The inference with the Bayesian linear model can be found in standard textbooks \citep{Denison:2002}. We assumed conjugate priors. These assumptions imply that the marginal prior distributions of the regression parameters are the multivariate t-distributions and the square of the standard deviation of the dependent variable error ($\sigma^2$) follows the inverse gamma distribution. The parameters of the marginal posterior distributions can be computed analytically and follow the same standard distributions as priors. 
 
 We used the following parameter values:
 \begin{itemize}  
  \item for noise of the dependent variable $B_\textrm{ver}$ , we set the parameters of the inverse gamma distribution to obtain the mode of the marginal prior distribution of $\sigma,$ which is about 145~G, which corresponds to the value estimated from the observed data,
  \item for noise of the dependent variable $B,$ we set the parameters of the inverse gamma distribution to obtain the mode of the marginal prior distribution of $\sigma,$ which is about 165~G, which corresponds to the value estimated from the observed data,
  \item for noise of the dependent variable $\gamma,$ we set the parameters of the inverse gamma distribution to obtain the mode of the marginal prior distribution of $\sigma,$ which is
 about 5 degrees, which corresponds to the value estimated from the observed data, 
  \item the prior distributions of the intercept values of $\gamma$, $B$, and $B_\textrm{ver}$ have a mean corresponding to the mean value of the observed data of each of the dependent variable, and the t-distribution is scaled by the standard deviation of the error of the dependent variables,
  \item the prior distributions of the slope values of $\gamma$, $B$, and $B_\textrm{ver}$ have a mean of zero to eliminate any preferences, and the distributions are scaled by the standard deviation,
  \item the prior distributions of the quadratic coefficients of $\gamma$, $B$, and $B_\textrm{ver}$ have a mean of zero to eliminate any preferences, and the distributions are scaled by the standard deviation.
 \end{itemize}

\begin{figure}[!t]
 \centering \includegraphics[width=0.97\linewidth]{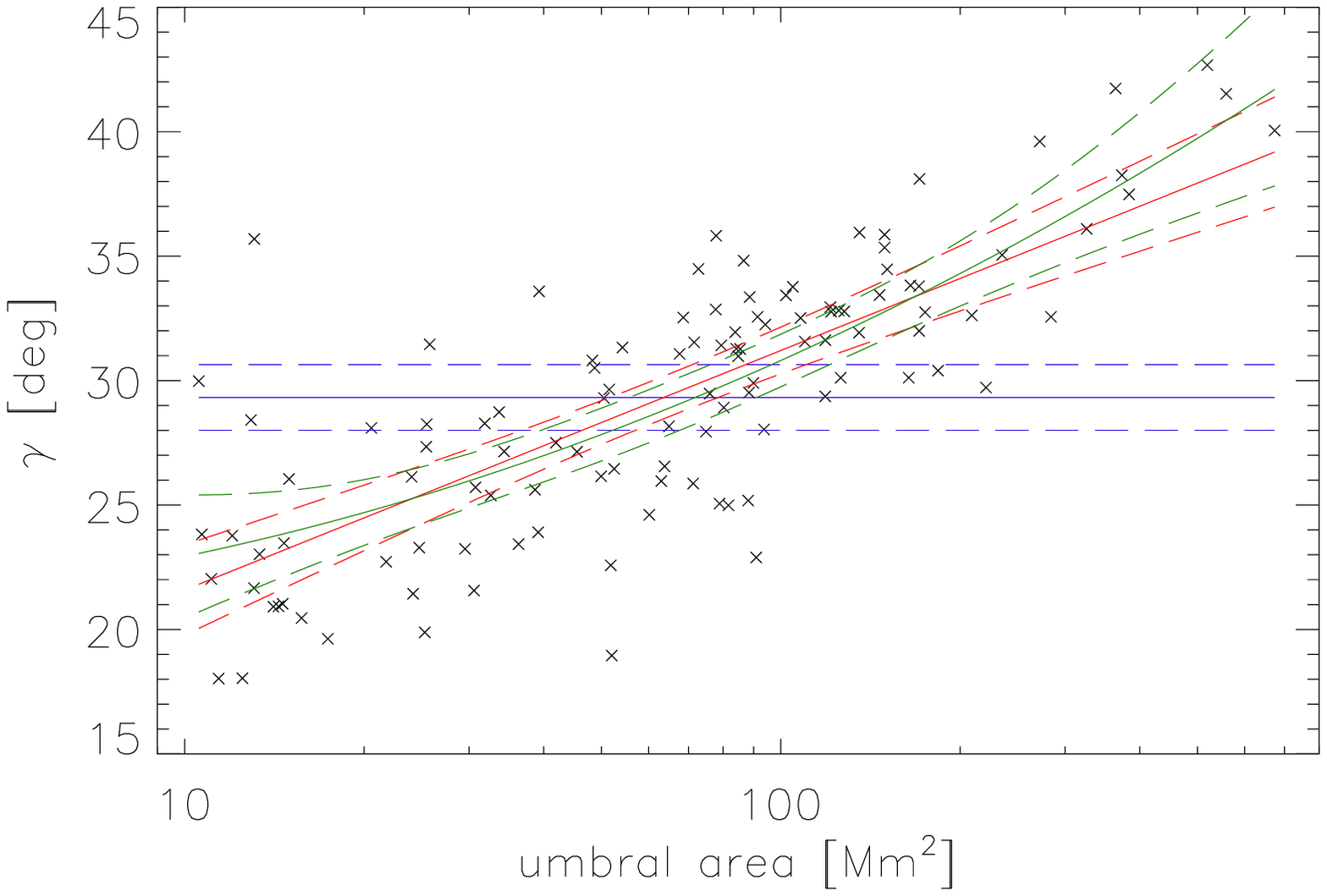}
 \caption{Scatter plot showing the dependence of magnetic field inclination on the umbra area. $\text{A black cross}$ marks the observed data points. The line colour denotes the model complexity of the Bayesian linear regression; blue shows a constant model, red a linear model, and green a quadratic model. The solid lines mark the most probable value estimated by the corresponding model,
and the dashed lines mark the 99\% confidence interval of the estimated model.}
 \label{SM_bayes_inc}
\end{figure}

\begin{figure}[!t]
 \centering \includegraphics[width=0.97\linewidth]{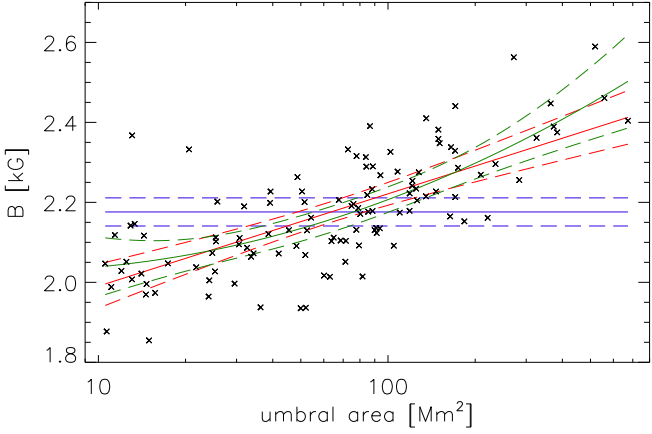}
 \caption{Same as Fig.~\ref{SM_bayes_inc}, but for the dependence of the magnetic field strength on the umbral area.}
 \label{SM_bayes_b}
\end{figure}

\begin{figure}[!t]
 \centering \includegraphics[width=0.97\linewidth]{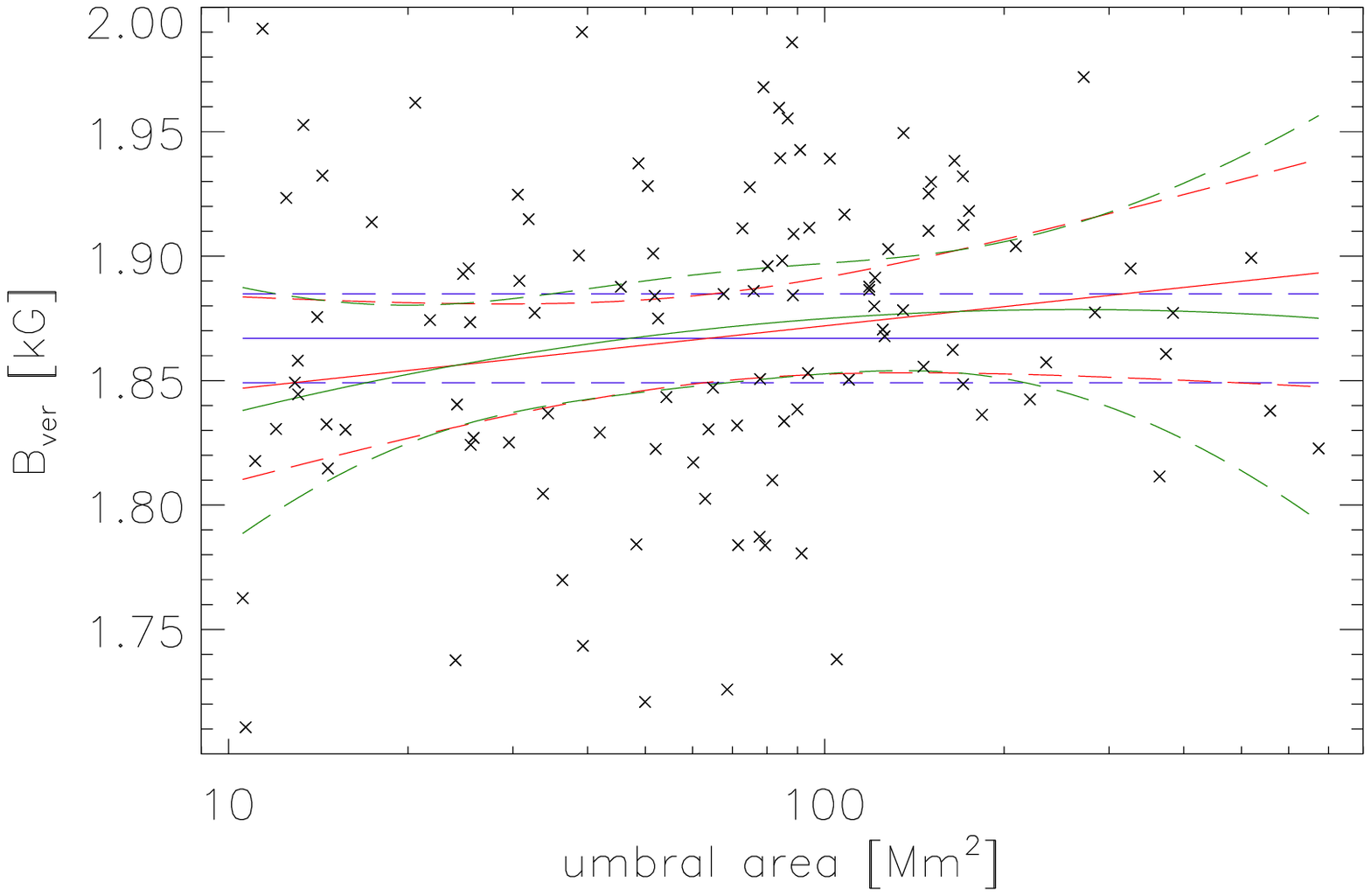}
 \caption{Same as Fig.~\ref{SM_bayes_inc}, but for the dependence of $B_\textrm{ver}$ on the umbral area.}
 \label{SM_bayes_bverarea}
\end{figure}

\begin{figure}[!t]
 \centering \includegraphics[width=0.97\linewidth]{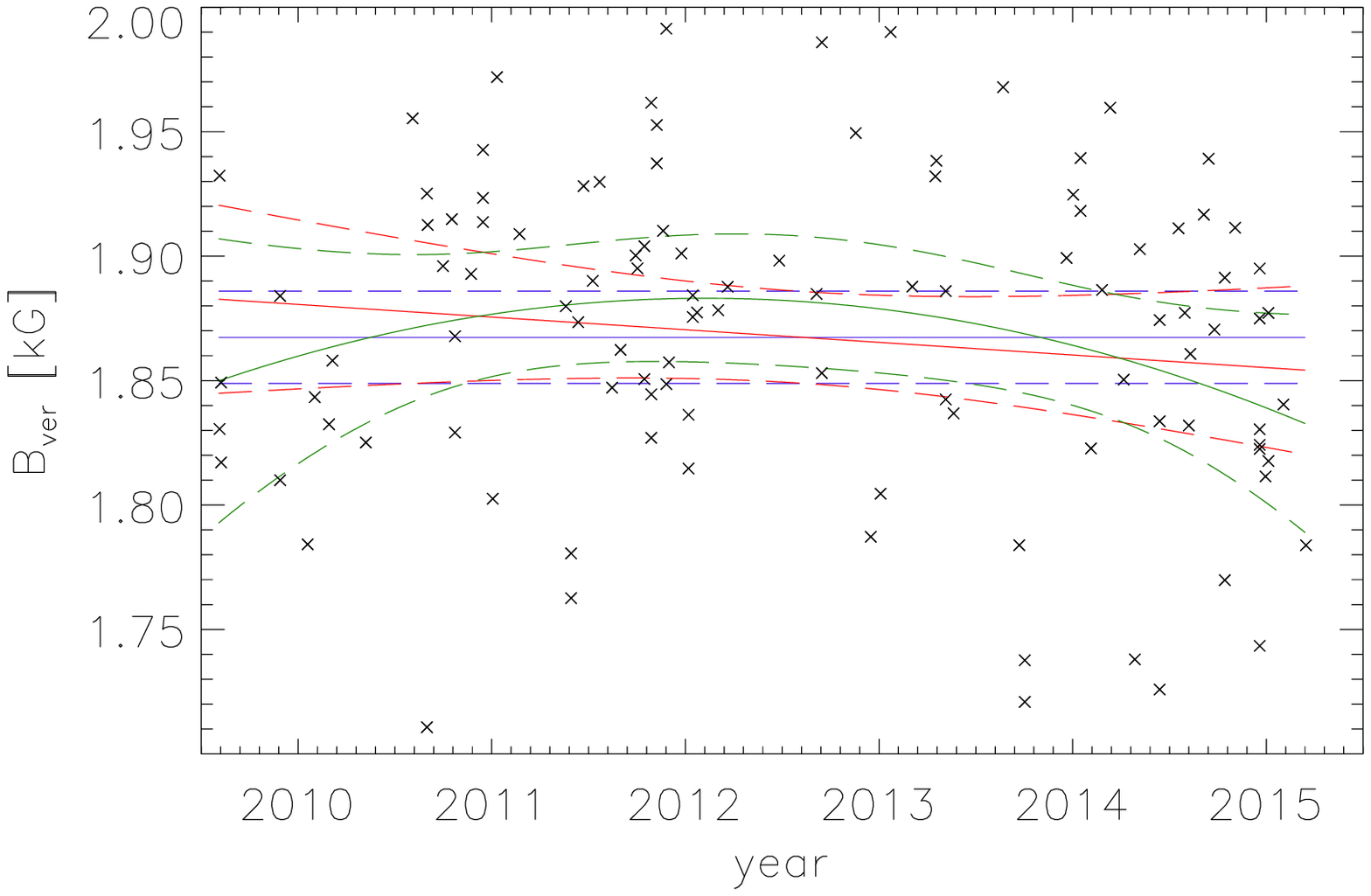}
 \caption{Same as Fig.~\ref{SM_bayes_inc}, but for the dependence of $B_\textrm{ver}$ on the phase of solar cycle~24.}
 \label{SM_bayes_bverdate}
\end{figure}

\begin{figure}[!t]
 \centering \includegraphics[width=0.97\linewidth]{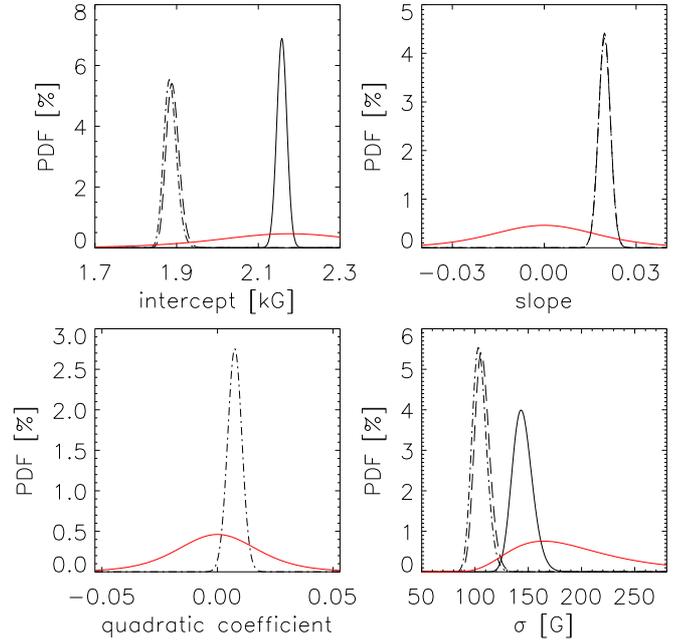}
 \caption{Prior (red lines) and posterior (black lines) distributions of the regression coefficients and standard deviation  of the error distribution for the dependence of magnetic field inclination on the logarithm of the area. The solid lines correspond to the constant model, the dashed lines to the linear model, and the dash-dotted lines to the quadratic model.}
 \label{SM_prior_inc}
\end{figure}

\begin{figure}[!t]
 \centering \includegraphics[width=0.97\linewidth]{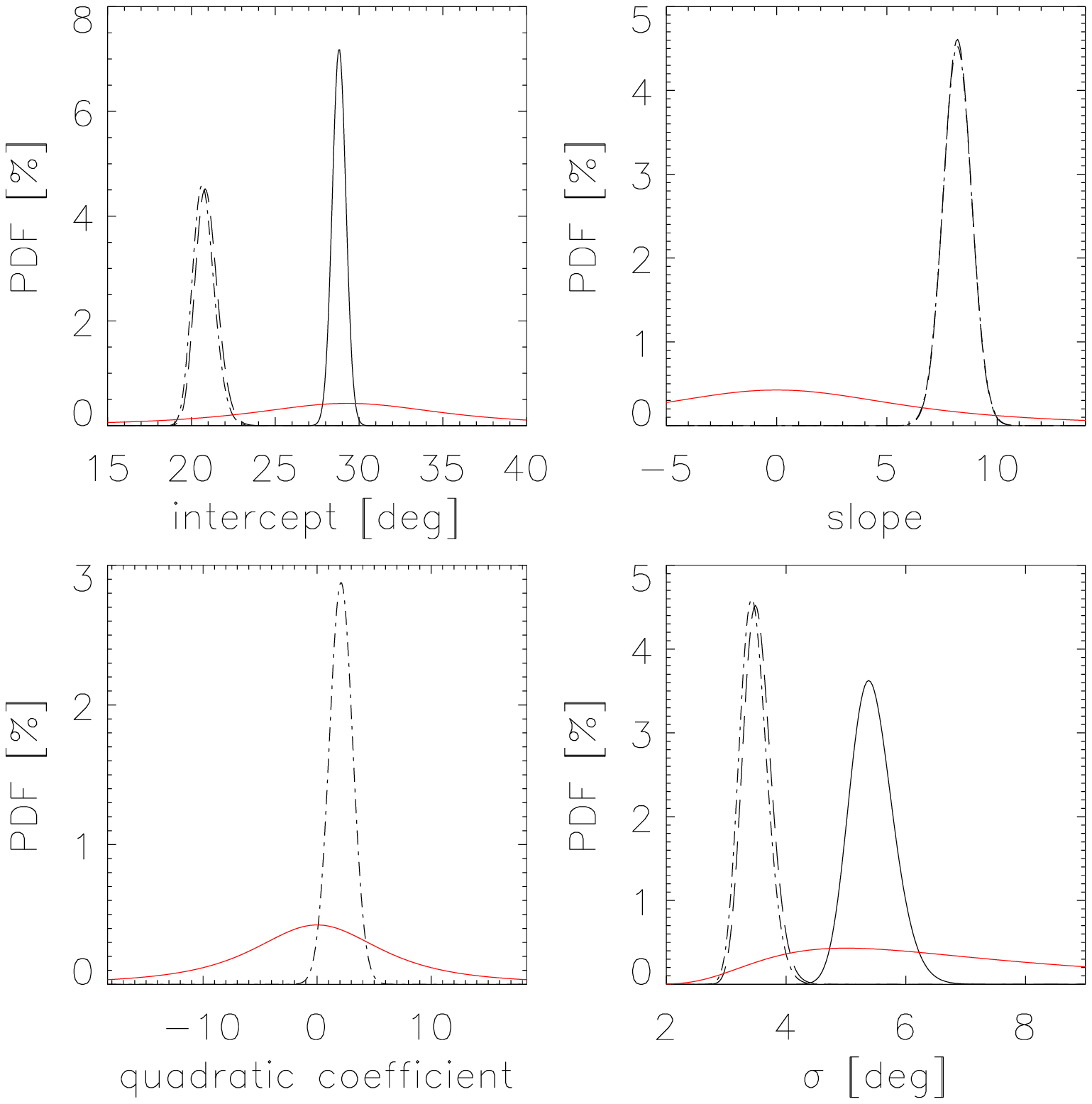}
 \caption{Same as Fig.~\ref{SM_prior_inc}, but for the dependence of B on the logarithm of the area.}
 \label{SM_prior_b}
\end{figure}

\begin{figure}[!t]
 \centering \includegraphics[width=0.97\linewidth]{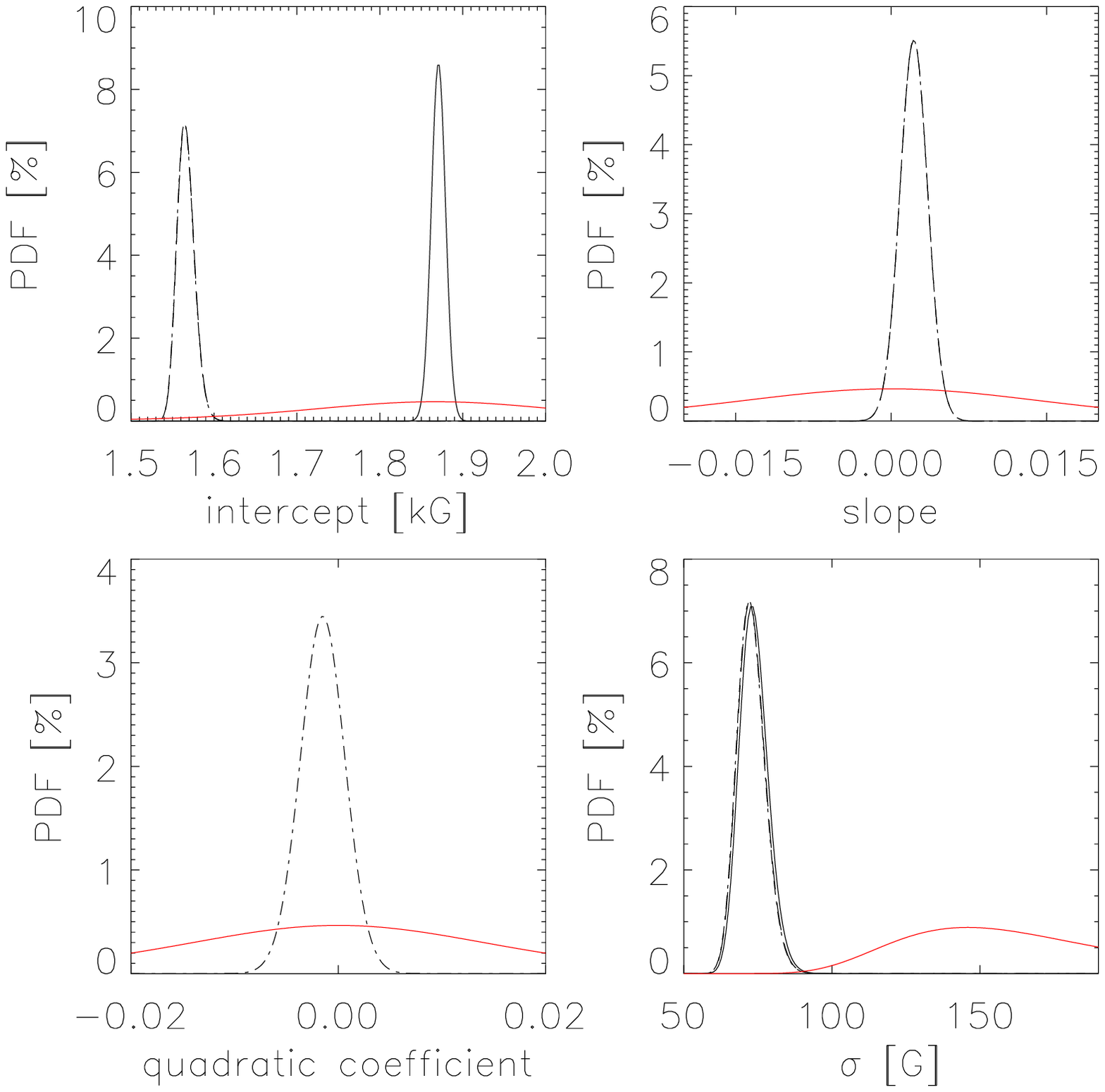}
 \caption{Same as Fig.~\ref{SM_prior_inc}, but for the dependence of $B_\textrm{ver}$ on the logarithm of the area.}
 \label{SM_prior_bverarea}
\end{figure}

\begin{figure}[!t]
 \centering \includegraphics[width=0.97\linewidth]{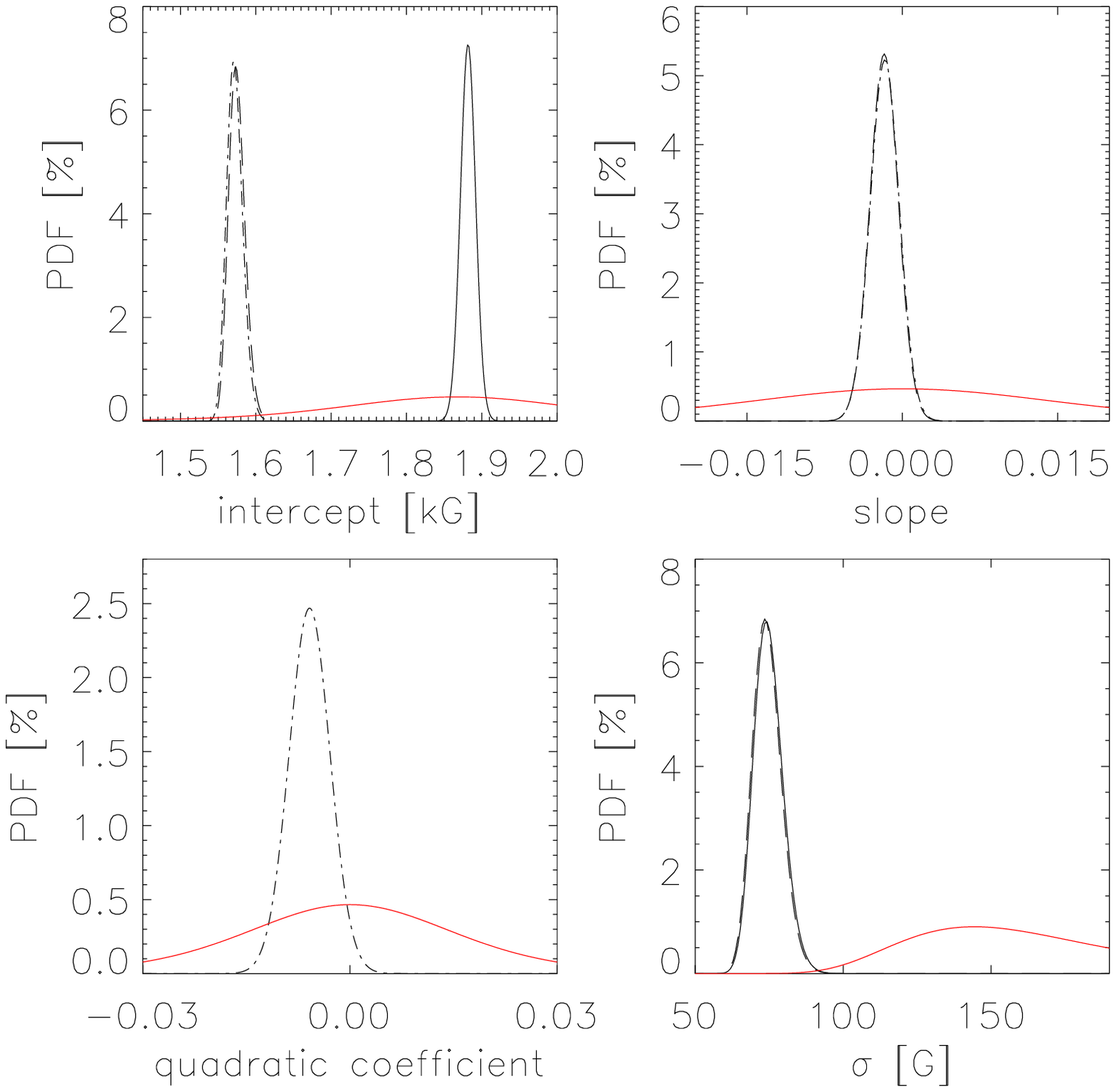}
 \caption{Same as Fig.~\ref{SM_prior_inc}, but for the dependence of $B_\textrm{ver}$ on the phase of solar cycle~24.}
 \label{SM_prior_bverdate}
\end{figure}

 The results of Bayesian linear regression are shown in Figs.~\ref{SM_bayes_inc}-\ref{SM_bayes_bverdate}. The most likely solutions and the 99\% confidence intervals presented in these figures are derived from the resulting posterior distributions that are shown in Figs.~\ref{SM_prior_inc}-\ref{SM_prior_bverdate}. The prior and posterior distributions of the regression parameters are presented using the scaled and centred explanatory variables.
 
 For every learned model $M,$ we computed the marginal log-likelihood of the model and compared the constant model ($M_0$), the linear model ($M_1$), and the quadratic model ($M_2$) using the Bayes factors (BF), which is defined as the ratio of marginal likelihoods of the two compared models. We summarise the results in Tables~\ref{bayes_inc}-\ref{bayes_Bverdate}. The hypothesis for each element of the matrix is in favour of the model associated with the row against the model associated with the column. A widely used interpretation \citep{Jeffreys:1961} of the strength of evidence for Bayes factor B is the following:
 \begin{itemize}
  \item $\mathrm{BF} \leq 0.1$ -- strong against,
  \item $0.1 < \mathrm{BF} \leq 1/3$ -- substantial against,
  \item $1/3 < \mathrm{BF} < 1$ -- barely worth mentioning against,
  \item $1 \leq \mathrm{BF} < 3$ -- barely worth mentioning for,
  \item $3 \leq \mathrm{BF} < 10$ -- substantial for,
  \item $10 \leq \mathrm{BF}$ -- strong for.  
 \end{itemize}

 \begin{table}
  \caption{Bayes factors for the magnetic field inclination as a function of the area logarithm.}
\label{bayes_inc}
\centering 
\begin{tabular}{c  c c c }
model & constant & linear & quadratic \\
\hline 
constant & 1.0000000 & 7.655851e-22 & 3.260399e-22 \\

linear & 1.306191e+21 & 1.0000000 & 4.258703e-01 \\

quadratic & 3.067109e+21 & 2.348133e+00 & 1.000000\\
\hline
\end{tabular}
 \end{table}
 
  \begin{table}
  \caption{Bayes factors for the magnetic field strength as a function of the area logarithm.}
\label{bayes_B}
\centering 
\begin{tabular}{c  c c c }
model & constant & linear & quadratic \\
\hline 
constant & 1.0000000 & 8.092029e-16 & 1.501854e-16 \\

linear & 1.235784e+15 & 1.0000000 & 1.855967e-01 \\

quadratic & 6.658439e+15 & 5.388028e+00 & 1.000000\\

\end{tabular}
 \end{table}

   \begin{table}
  \caption{Bayes factors for the $B_\textrm{ver}$ as a function of the area logarithm.}
\label{bayes_Bverarea}
\centering 
\begin{tabular}{c c c c }
model & constant & linear & quadratic \\
\hline 
constant & 1.0000000 & 1.4529998 & 3.824566 \\

linear & 0.6882313 & 1.0000000 & 2.632186 \\

quadratic & 0.2614676 & 0.3799123 & 1.000000\\
\hline
\end{tabular}
 \end{table}

    \begin{table}
  \caption{Bayes factors for the $B_\textrm{ver}$ as a function of the phase of solar cycle~24.}
\label{bayes_Bverdate}
\centering 
\begin{tabular}{c c c c}
model & constant & linear & quadratic \\
\hline 
constant & 1.0000000 & 2.541874 & 0.9061921 \\

linear & 0.3934106 & 1.0000000 & 0.3565056 \\

quadratic & 1.1035187 & 2.805005 & 1.000000\\
\hline
\end{tabular}
 \end{table}
 
 We can draw the following conclusions from Tables~\ref{bayes_inc}-\ref{bayes_Bverdate} and Figs.~\ref{SM_bayes_inc}-\ref{SM_prior_bverdate}:
 \begin{itemize}
  \item There is strong evidence in the data that the magnetic field inclination  is not constant and is a function of the logarithm of the area. The resulting confidence intervals do not allow us to distinguish if the dependence of $\gamma$ on the logarithm of the area is linear or quadratic.
  \item There is strong evidence in the data that the strength of the magnetic field  is not constant and is a function of the logarithm of the area. The resulting confidence intervals do not allow us to distinguish if the dependence of $B$ on the logarithm of the area is linear or quadratic.
  \item There is no substantial evidence in the data that the vertical component of the magnetic field  is a linear or a quadratic function of the logarithm of the area. The constant solution is well within the confidence intervals of the linear and quadratic solutions.
  \item There is no substantial evidence in the data that the vertical component of the magnetic field  is a linear or a quadratic function of the date. The constant solution is well within the confidence intervals of the linear and quadratic solutions.
 \end{itemize}

Figures~\ref{SM_bayes_inc}-\ref{SM_bayes_bverdate} show that the 99\% confidence intervals of the estimated models are quite narrow compared to the spread of the measured values of dependent variables. This can be explained by the fact that the Bayesian analysis allows splitting the overall uncertainty between the uncertainty of model parameters (which we see in these figures) and the error represented by the value of the standard deviation (see the bottom right plots in Figs.~\ref{SM_prior_inc}-\ref{SM_prior_bverdate}).
 
 \subsection{Standard linear regression}
 
 The goal of the standard linear regression \citep{Weisberg:2005} is to estimate the vector of the linear regression coefficients $\beta$ from the measured data. We estimated the coefficients by solving the linear least-squares problem using the QR factorization. Each coefficient is a normally distributed random variable. For each coefficient we report
 \begin{itemize}
  \item its value,
  \item the standard error of each coefficient, which is the standard deviation of its distribution. It measures the uncertainty in the estimate of the coefficient,
  \item its p-value,  which is the probability of achieving the same or a higher value of the t-statistics if the null hypothesis were true. The null hypothesis is that the corresponding value of $\beta$ is zero.  If the p-value was higher than 0.01, we did not reject the null hypothesis. The p-value was computed for the coefficients learned from the scaled and centred data.
 \end{itemize}

 For each model we also report the p-value, that is, the probability of achieving a value of the t-statistics as high or higher if the null hypothesis were true, where the null hypothesis is the intercept-only model. If the p-value was higher than 0.01, we did not reject the null hypothesis. With this p-value, the probability of incorrectly rejecting a true null hypothesis is typically close to 15\% \citep{Sellke:2001}.
 
\begin{figure}[!t]
 \centering \includegraphics[width=0.97\linewidth]{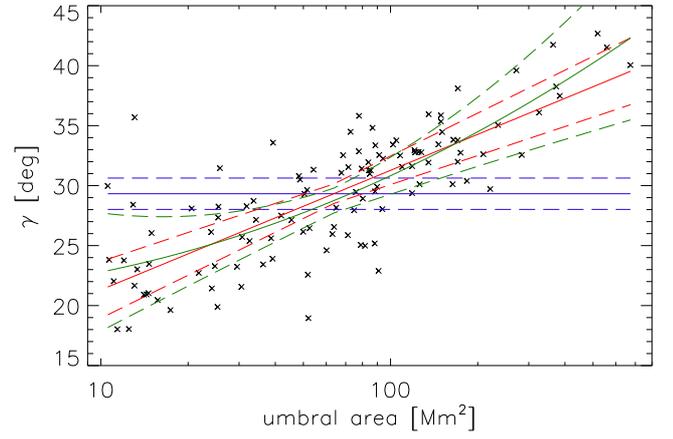}
 \caption{Scatter plot showing the dependence of the magnetic field inclination on the umbra area. Black crosses mark the observed data points. The line colour denotes the model complexity of the standard linear regression; blue shows a constant model, red a linear model, and green a quadratic model. The solid lines mark the the best fit, and the dashed lines mark the 99\% confidence interval of the estimated model.}
 \label{SM_classical_inc}
\end{figure}

\begin{figure}[!t]
 \centering \includegraphics[width=0.97\linewidth]{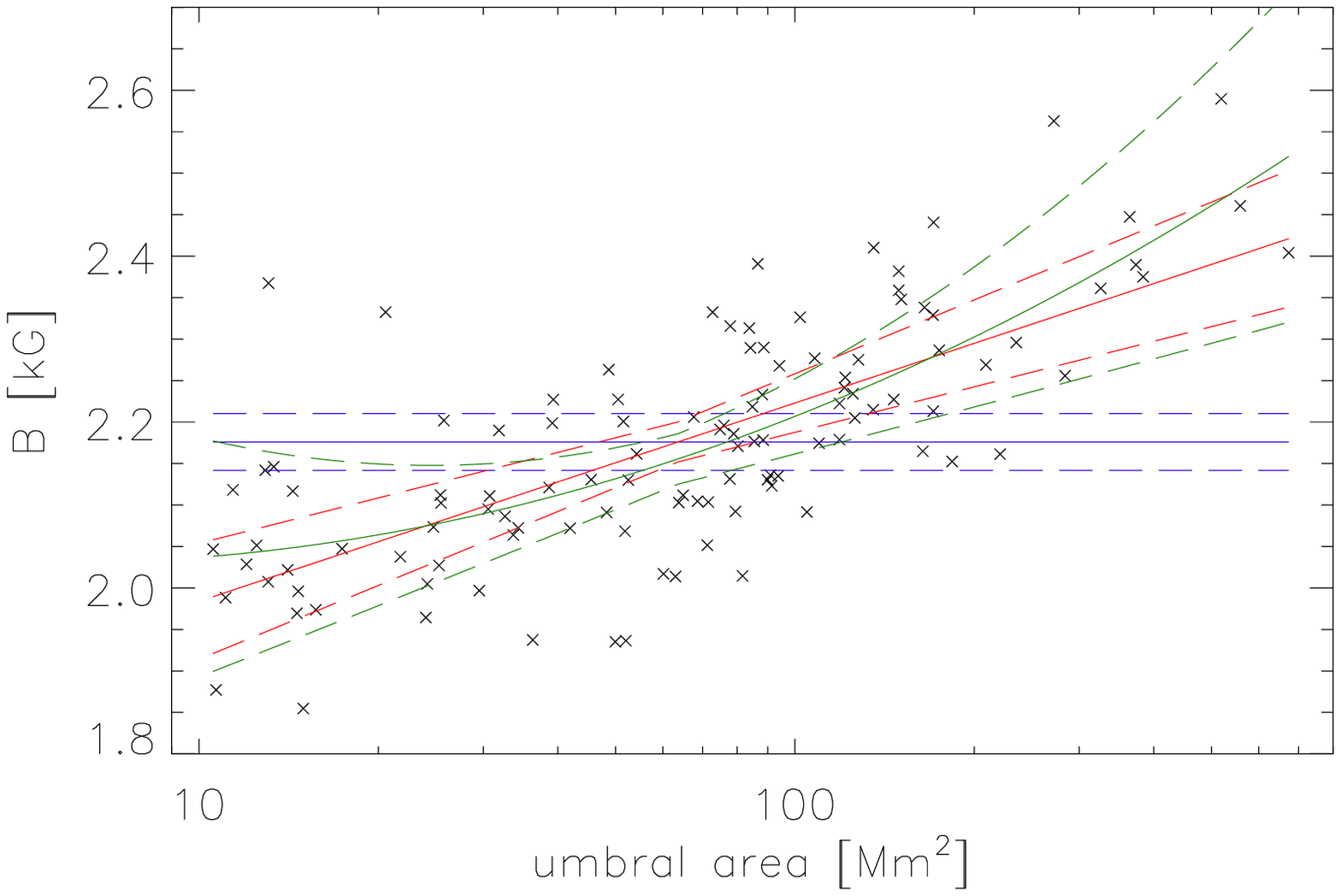}
 \caption{Same  as Fig.~\ref{SM_classical_inc}, but for the dependence of $B$ on the logarithm of the umbral area.}
 \label{SM_classical_b}
\end{figure}

\begin{figure}[!t]
 \centering \includegraphics[width=0.97\linewidth]{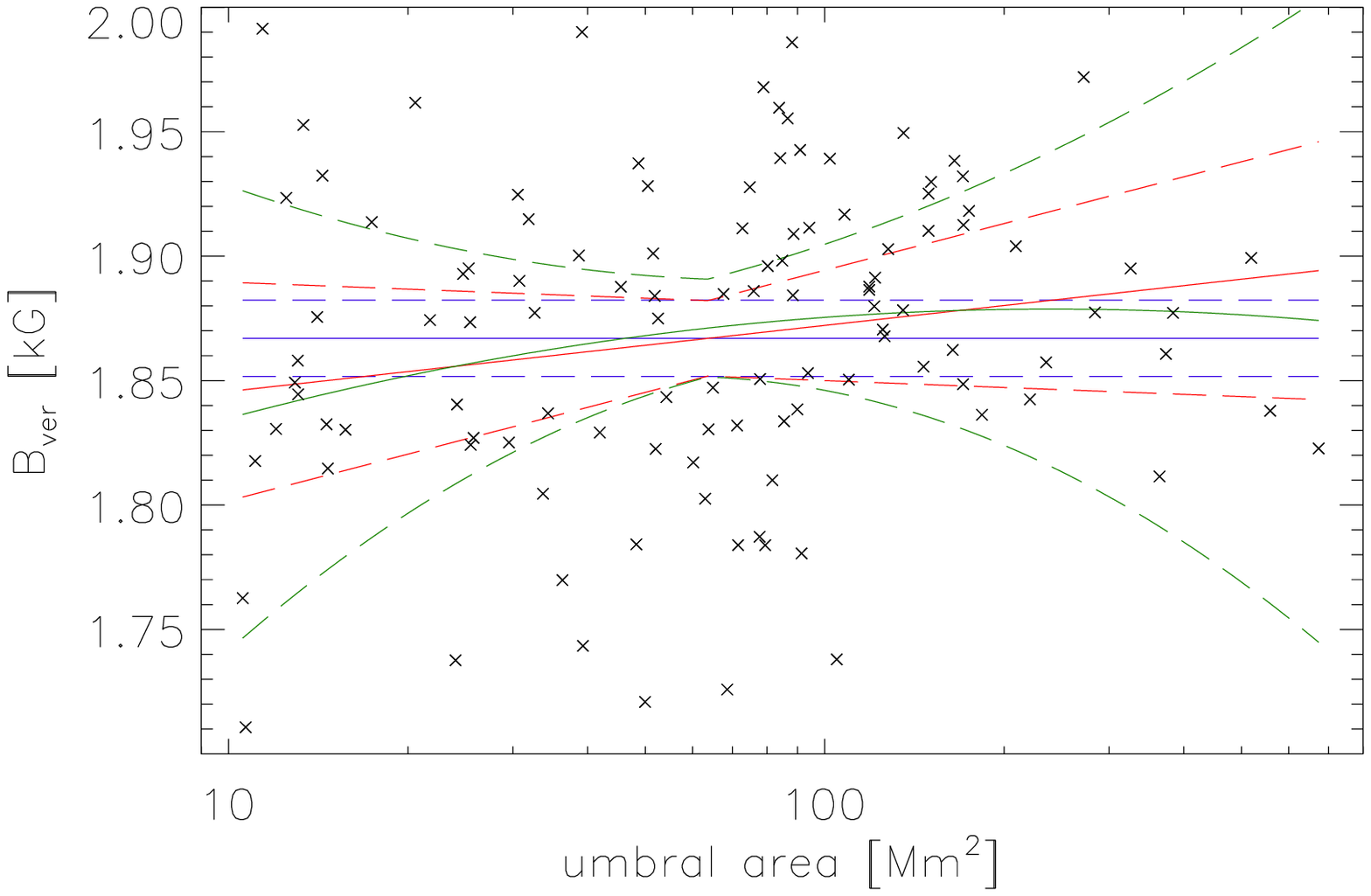}
 \caption{Same  as Fig.~\ref{SM_classical_inc}, but for the dependence of $B_\textrm{ver}$ on the logarithm of the umbral area.}
 \label{SM_classical_bverarea}
\end{figure}

\begin{figure}[!t]
 \centering \includegraphics[width=0.97\linewidth]{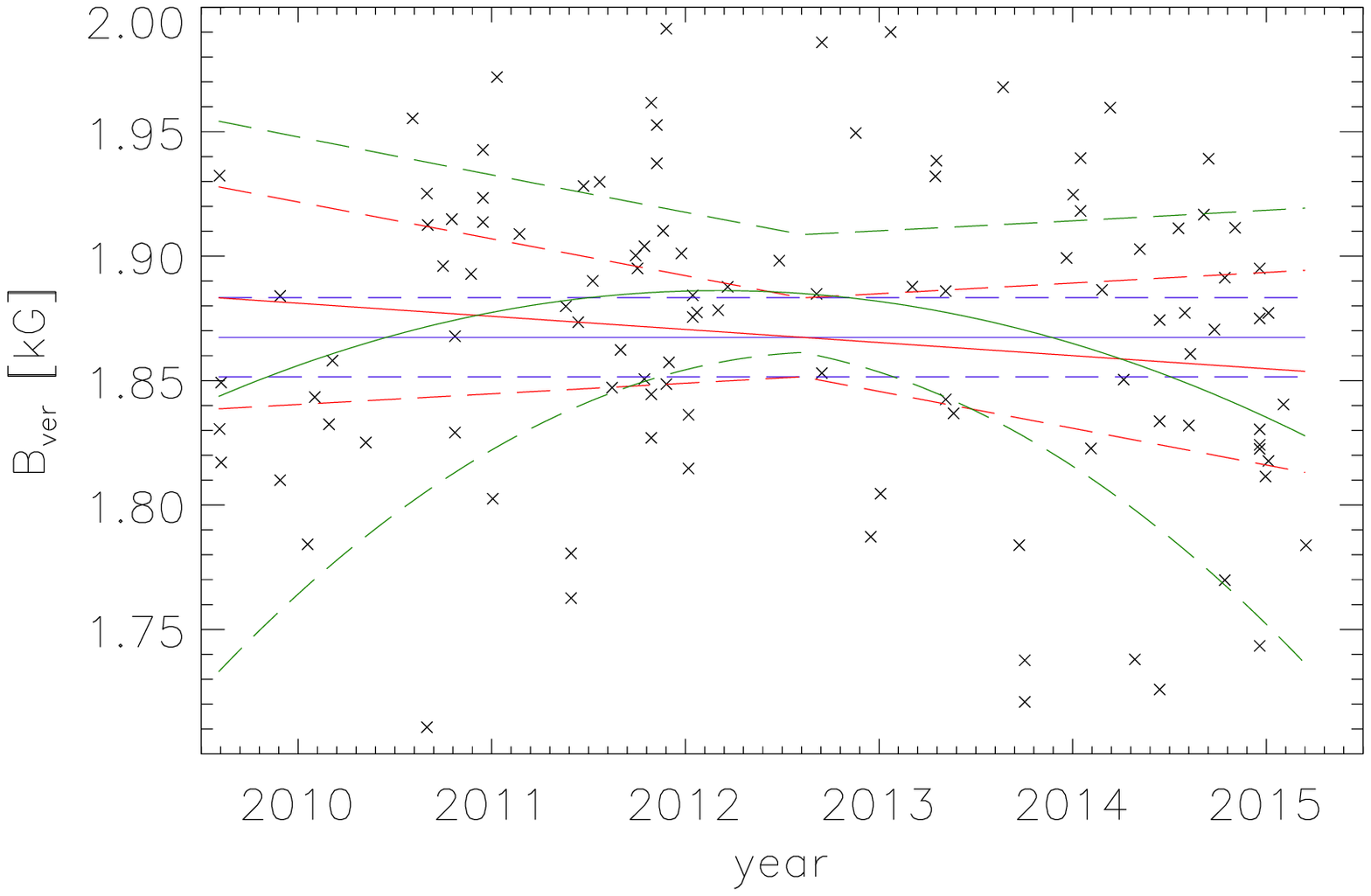}
 \caption{Same  as Fig.~\ref{SM_classical_inc}, but for the dependence of $B_\textrm{ver}$ on the phase of solar cycle~24.}
 \label{SM_classical_bverdate}
\end{figure}

 The resulting values of standard linear regression are listed in Tables~\ref{clasic_inc}-\ref{clasic_bverdate}. These values are used to plot the best fit and the 99\% confidence intervals in Figs.~\ref{SM_classical_inc}-\ref{SM_classical_bverdate}. 
 
 The results of the standard linear regression are in full agreement with the results of the Bayesian linear regression:
 \begin{itemize}
  \item There is strong evidence in the data that the magnetic field inclination is not constant and is a function of the logarithm of the area as the p-values of both linear and quadratic models are extremely low and we can reject the constant model.
  \item There is strong evidence in the data that the strength of the magnetic field  is not constant and is a function of the logarithm of the area as the p-values of both linear and quadratic models are extremely low and we can reject the constant model.
  \item There is no substantial evidence in the data that the vertical component of the magnetic field  is a linear or a quadratic function of the logarithm of the area. The p-values of the more complex models are above 0.01, that is, we cannot reject the constant model.
  \item There is no substantial evidence in the data that the vertical component of the magnetic field  is a linear or a quadratic function of the date. The p-values of the linear and quadratic models are above 0.01, meaning that we cannot reject the constant model.
 \end{itemize}

\begin{table}
\caption{Model parameters for $\gamma$ as a function of the area
logarithm.}
\label{clasic_inc}
\centering 
\begin{tabular}{l c c c}
 
model & constant & linear & quadratic \\
\hline
intercept & 29.3 deg & 11.6 deg & 21.1 deg \\
slope &  &  9.96 & -1.51 \\
quadratic coeff. &  &  & 3.18 \\
\hline
$\sigma$-intercept & 0.5 deg & 1.4 deg & 4.8 deg\\
$\sigma$-slope &  & 0.75 & 5.19\\
$\sigma$-quadratic coeff. &  &  & 1.42 \\
\hline
p-value intercept & $< $2e-16 & $< $2e-16 & $< $2e-16 \\
p-value slope &  & $< $2e-16 & $< $2e-16 \\
p-value quadratic coeff. &  &  & 0.027 \\
\hline
model p-value & NA & $< $2.2e-16 & $< $2.2e-16 \\
\hline
\end{tabular}
\end{table}
 
 \begin{table}
\caption{Model parameters for $B$ as a function of the area logarithm.}
\label{clasic_b}
\centering 
\begin{tabular}{l c c c }

model & constant & linear & quadratic \\
\hline
intercept & 2176~G deg & 1745 G & 2095 G \\
slope &  &  0.024 & -0.017 \\
quadratic coeff. &  &  & 0.011 \\
\hline
$\sigma$-intercept & 13 G & 41 G & 142 G\\
$\sigma$-slope &  & 0.002 & 0.015\\
$\sigma$-quadratic coeff. &  &  & 0.004 \\
\hline
p-value intercept & $< $2e-16 & $< $2e-16 & $< $2e-16 \\
p-value slope &  & $< $2e-16 & $< $2e-16 \\
p-value quadratic coeff. &  &  & 0.007 \\
\hline
model p-value & NA & $< $2.2e-16 & $< $2.2e-16 \\
\hline
\end{tabular}
\end{table}

\begin{table}
\caption{Model parameters for $B_\textrm{ver}$ as a function of the area logarithm.}
\label{clasic_bverarea}
\centering 
\begin{tabular}{l c c c }

model & constant & linear & quadratic \\
\hline 
intercept & 1867~G deg & 1819 G & 1748 G \\
slope &  &  0.00266 & 0.01095 \\
quadratic coeff. &  &  & -0.0023 \\
\hline
$\sigma$-intercept & 6 G & 26 G & 92 G\\
$\sigma$-slope &  & 0.00139 & 0.00983 \\
$\sigma$-quadratic coeff. &  &  & 0.0027 \\
\hline
p-value intercept & $< $2e-16 & $< $2e-16 & $< $2e-16 \\
p-value slope &  & 0.0583 & 0.059 \\
p-value quadratic coeff. &  &  & 0.396 \\
\hline
model p-value & NA & 0.0583 & 0.117 \\
\hline
\end{tabular}
\end{table}

\begin{table}
\caption{Model parameters for $B_\textrm{ver}$ as a function of the phase of solar cycle~24.}
\label{clasic_bverdate}
\centering 
\begin{tabular}{l c c c }

model & constant & linear & quadratic \\
\hline 
intercept & 1867~G deg & 12.5 kG & -25790 kG \\
slope &  &  -0.0005 & 2.564 \\
quadratic coeff. &  &  & -0.0006 \\
\hline
$\sigma$-intercept & 6 G & 7000 kG & 10190 kG\\
$\sigma$-slope &  & 0.0004 & 1.013 \\
$\sigma$-quadratic coeff. &  &  & 0.0003 \\
\hline
p-value intercept & $< $2e-16 & $< $2e-16 & $< $2e-16 \\
p-value slope &  & 0.158 & 0.132 \\
p-value quadratic coeff. &  &  & 0.013 \\
\hline
model p-value & NA & 0.158 & 0.017 \\
\hline
\end{tabular}
\end{table}

\end{appendix}

\end{document}